\documentclass[twocolumn]{jpsj2} 

\newcommand{\bcite}[1]{\cite{#1}}

\newcommand{\Ham}{{\cal H}}

\newcommand{\vect}[1]{{\mbox{\boldmath $#1$}}}

\newcommand{\figlabel}[1]{\label{fig:#1}}

\newcommand{\Eq}[1]{(\ref{eq:#1})}
\newcommand{\Fig}[1]{Fig.\ \ref{fig:#1}}

\newcommand{\deffig}[4]{
\begin{figure}[tb]
  \begin{center}
  \includegraphics[width=#3 \textwidth]{Fig/#2}
  \end{center}
  \caption{ \figlabel{#1} #4}
\end{figure}
}

\newcommand{\KCC}{KCuCl$_3$}
\newcommand{\TCC}{TlCuCl$_3$}

\title{
  Critical Properties of 
  Condensation of Field-Induced Triplet Quasiparticles
}
\author{
  Naoki \textsc{KAWASHIMA}
  \thanks{E-mail address: kawashima@issp.u-tokyo.ac.jp}
}
\inst{
          Department of Physics, 
          Tokyo Metropolitan University, 
          Hachioji, Tokyo 192-0357, Japan; \\
          Institute for Solid State Physics, 
          University of Tokyo, Kashiwa, Chiba 277-8581, Japan
}
\abst{
A review on the field-induced magnetic ordering is given, 
together with some results of a quantum Monte Carlo 
simulation focused on the critical behevior near the
quantum critical point.
}
\kword{quantum spin system, Heisenberg model, quantum Monte Carlo,
XY model, XXZ model, cluster algorithm, loop algorithm,
worm algorithm, directed-loop algorithm}

\begin{document}
\maketitle

\section{Introduction}

In this article we discuss phenomena 
observed in systems with a singlet ground state
with triplet excitations separated by a finite spin gap.
The focus of researchers' attention has been 
on the magnetic phase transition
induced by a magnetic field and/or a pressure.
While the field- (or pressure-) induced magnetic ordering 
were observed in various materials,
the character of the phase transition, 
in particular the one at the quantum critical point (QCP)
seems to be insensitive 
to the the nature of the ground state and
to the origin of the gap.
In one class of materials, localized magnetic spins are bound
strongly in pairs to form singlet dimers, and the inter-dimer
interaction is not strong enough to induce any long-range
magnetic ordering.
In these materials, the excited states consist of 
excitations of quasiparticles,
recently often referred to as {\it tripletons} or {\it triplons}.
They originate from the triplet states at each site.
The magnetic ordering in the dimer systems with a sufficiently
strong magnetic field was suggested based on a mean-field theory.
\cite{TachikiY1970a,TachikiY1970b}
Later, two possibilities were suggested\cite{Rice2002}
depending on the relative strength of the 
repulsive interaction among the quasiparticles
and their kinetic energy.
If the kinetic term dominates,
the system may undergo a condensation transition
as suggested by the mean-field theory\cite{TachikiY1970a,TachikiY1970b}
and later also by the Hartree-Fock (HF) theory with the bosonic 
representation.\cite{NikuniOOT2000}
On the other hand, when the repulsive force dominates, 
the ordered state
is the super lattice where the quasiparticles are
located periodically to form a lattice with a larger
unit cell than that of the original lattice, which gives rise
to magnetization plateaus in the magnetization-field curve.
This scenario is realized in materials for which
the effective dimer-dimer couplings vanish 
due to the cancellation among the exchange couplings
and higher order interactions play essential roles.
\cite{Fukumoto2001} 
A well-known example is SrCu$_2$(BO$_3$)$_2$.
\cite{KageyamaETAL1999,MiyaharaU2000,KodamaETAL2002}
In the rest of the present article, however, 
we only discuss the first case,
i.e., the case where the kinetic-term is dominant.

\section{Early Experiments and Mean-Field Theory}

A long lasting flow of publications on the field-induced magnetic ordering
was initiated by a theoretical work\cite{TachikiY1970a,TachikiY1970b},
which was inspired by experiments\cite{AmayaTYAH1969,HasedaETAL1970} 
on a quasi one-dimensional magnetic compound 
Cu(NO$_3$)$_2\cdot 2.5$ H$_2$O.
The zero-field susceptibility measurement\cite{BergerFS1963}
had been suggesting that the system's main magnetic structure 
consisting of isolated dimers.
Heat capacity measurements\cite{FriedbergR1968}
and magnetization measurements\cite{MyersBF1969}
in the presence of a magnetic field were also carried out.
However, no clear indication of a phase transition was observed.
In retrospect, this was simply because the temperature was not
low enough or the magnetic field was not strong enough to bring 
the system into an ordered state.
Later, an unusual magnetocaloric behavior was observed\cite{AmayaTYAH1969},
which seemed to indicate some sort of phase transition.
Subsequently, the existence of the transition was confirmed by an
anomaly in the adiabatic magnetization curve\cite{HasedaETAL1970}
at 3-4 T below 1 K.

In order to understand this phenomena,
a model system that consists of dimers of spins
was considered.\cite{TachikiY1970a,TachikiY1970b}
The building block of the model is a pair of $S=1/2$ spins 
coupled to each other antiferromagnetically.
When a magnetic field is applied to the system,
the degeneracy among the triplet states
is lifted and the lowest branch comes down
to meet the singlet state.
To be concrete, the spin Hamiltonian considered was
\begin{equation}
  \Ham =     \sum_i \hat J \vect{s}_{i1}\cdot\vect{s}_{i2}
           + \sum_{(ij)}\sum_{\alpha,\beta} \hat J'_{i\alpha,j\beta} 
             \vect{s}_{i\alpha}\cdot\vect{s}_{j\beta}
           - \sum_{i\alpha} \hat H s^z_{i\alpha}.
  \label{eq:OriginalHamiltonian}
\end{equation}
Here $\hat J>0$ is the antiferromagnetic coupling constant that binds
a pair of spins, $i$ and $j$ the indices of dimers, and 
$\alpha$ and $\beta$ the specifiers of individual spins in a dimer.
The symbol $\vect{s}_{i\alpha}$ stands for an $S=1/2$
spin operator.
The dominant couplings are the intra-dimer couplings
represented by $\hat J$.
While the compound Cu(NO$_3$)$_2\cdot 2.5$ H$_2$O seemed
to have a quasi-one-dimensional lattice structure,
\cite{BonnerFKM1970}
any specific structure was assumed
in the mean-field theory because it is not relevant 
in the approximation.
In fact, as long as the lattice is three dimensional,
its structure does not affect
asymptotic behaviors near the criticality
even in more elaborate treatments.
Therefore, in most of the following discussion
we only assume that the lattice is three dimensional. 

The above Hamiltonian can be further simplified, taking into
account the fact that we can neglect 
local high energy levels, when the inter-dimer couplings are 
sufficiently small and can be treated as a perturbation.
Only two lower levels, the singlet state and one of the triplet
states are retained.
These two levels can be regarded as the up and down 
states of an effective $S=1/2$ spin.
Then we have the following effective Hamiltonian:\cite{TachikiY1970a}
\begin{equation}
  \Ham =
           \sum_{(ij)} \tilde J_{ij}
           \left( S_i^x S_j^x + S_i^y S_j^y + \frac12 S_i^z S_j^z \right)
         - \sum_i \tilde H_i S_i^z
  \label{eq:ReducedHamiltonian}
\end{equation}
where $S_i^{\mu}$ is an operator representing the effective spin.
The coupling $\tilde J_{ij}$ and $\tilde H_i$ can be expressed as simple
linear combinations of $\hat J$, $\hat H$ and $\hat J'_{i\alpha,j\beta}$.

We here consider the case where the effective Hamiltonian is 
translation invariant and non-frustrated.
The former is the case with most of materials unless 
impurities affect the observations.
The latter assumption does not strictly hold.
However, in many important cases, the frustration is weak,
at least not strong enough to alter the phase diagram structure
or the critical behavior qualitatively.
With these assumptions, we obtain an $S=1/2$ $XXZ$ model 
with easy plane anisotropy.
The mean-field approximation was applied to this model, yielding
\cite{TachikiY1970a}
$
  \Ham_{\rm mf} =
           \sum_{(ij)} \tilde J_{ij}
           \left( S_i^x m^x + \frac12 S_i^z m^z \right)
         - \sum_i \tilde H_i m^z.
  \label{eq:MeanFieldHamiltonian}
$
By solving the self-consistent equation in terms of $m^x$ and $m^z$,
a symmetry-breaking solution was obtained, from which 
the presence of a field-induced magnetic ordering in real materials
was suggested.

However, in subsequent works
it was found that the anomalous behavior observed in 
the adiabatic magnetization curve\cite{HasedaETAL1970}
of Cu(NO$_3$)$_2\cdot 2.5$ H$_2$O
should be attributed to the development of a short-range 
order along the chain, not a true long-range order.
By using the exact solution\cite{Katsura1962,Katsura1963}
to the $S=1/2$ $XY$ model in one dimension,
it was shown\cite{TachikiYM1970a,TachikiYM1970b}
that a purely one-dimensional model can produce
the adiabatic magnetization curve similar to the
one observed in the experiments.
Later, this result was confirmed by a more
extensive investigation.\cite{BonnerFKMB1983} 
An observation of the genuine long-range magnetic
ordering had to wait till the NMR experiment,
\cite{vanTolHP1971}
which revealed that the emergence of the long range order 
takes place in two stages; first the development of 
a short range order along the
chain at an intermediate temperature
and then the three-dimensional long range order 
at a lower temperature.
The heat capacity measurement\cite{vanTolDP1973} showed 
these characteristics more clearly in the form of two peaks;
a broad heat capacity peak at a high temperature and a sharper 
one at a low temperature.
The low-temperature one is identified as the transition 
discussed in the preceding theoretical works\cite{TachikiY1970a,TachikiY1970b}
based on the mean-field theory.
Subsequently, the magnetic structure in the ordered phase was determined
through a neutron diffraction experiment\cite{EckertCSFK1979} and 
an NMR experiment\cite{DiederixGHKP1978};
weakly coupled magnetic chains with each chain having
alternating coupling constants.

It should be pointed out here that the field-induced magnetic 
ordering itself is a rather generic phenomenon and not restricted 
to the dimer systems.
For example, a theoretical result similar to 
the one mentioned above
was obtained for an $S=1$ (not necessarily dimer) 
spin system with an uniaxial spin anisotropy\cite{TsunetoM1971}.
Due to the anisotropy, the single-site ground state is the
$S_z=0$ state whereas the $S_z = \pm 1$ states are degenerate
excited states. 
The mechanism of the magnetic ordering induced by a magnetic field
is quite analogous to that in the dimer system.
A heat capacity measurement was carried out
\cite{AlgraBDJC1977} on this type of material.
The compound was Ni(C$_5$H$_5$NO)$_6$(ClO$_4$)$_2$, in which
each Ni$^{2+}$ ion carries an $S=1$ spin.
The spins are influenced by the uniaxial spin anisotropy 
of which the magnitude is greater than the spin-spin exchange couplings.
A cusp-like peak was observed in the temperature dependence of
the heat capacity, and the parabolic phase diagram in the $H$-$T$
plane was drawn. 
There are several other reports on Ni based compounds.
A magnetization measurement\cite{PaduanFilhoCJC1981}
was performed for NiCl$_2\cdot$SC(NH$_2$)$_2$.
Later, heat capacity measurements
on Ni(C$_5$H$_14$N$_2$)$_2$N$_3$(ClO$_4$)
\cite{HondaETAL1997}
and Ni(C$_5$H$_14$N$_2$)$_2$N$_3$(PF$_6$)
\cite{HondaAK1998}
clearly showed a lambda peak in the heat capacity.
Compounds with more complicated spin anisotropies
were discussed within the mean-field theory
\cite{IshiiA1990} and the spin-wave theory.\cite{AoyamaI1992}
Recently, the rare-earth compound PrOs$_4$Sb$_{12}$ was studied 
through the heat capacity measurement\cite{AokiETAL2002},
and the neutron diffraction.\cite{KohgiETAL2003}
The level crossing scheme of this material 
was theoretically clarified.\cite{ShiinaA2004}

The Cr-based compound Cs$_3$Cr$_2$Br$_9$ may belong to
another category of materials that show a field-induced 
magnetic ordering.
This is a dimer system in which spins are bound 
pairwise antiferromagnetically to form dimers.
The Cr$^{3+}$ ions carry $S=3/2$ spins.
What makes this compound unique is the fact that
the ions form frustrated triangular lattices stacked 
on top of each other.
Due to the frustration,
the magnetic ordering in this compound seems to be more complicated
than a mere N\'{e}el state.
A cusp in the temperature dependence of the susceptibility was observed
\cite{LeuenbergerGHD1985,LeuenbergerGFK1985} in this compound
although the nature of the ordered phase was not clarified.
Corresponding to the four possible values of dimer angular momentum,
$S=0,1,2,3$, four plateaus were observed in the field dependence of
magnetization at low temperatures.\cite{AjiroKISG1989}
Quite recently elastic and inelastic neutron scattering experiments
\cite{GreinerETAL2004}
suggested that the gap does not close between two subsequent plateaus,
in contrast to the other materials mentioned above.
Theoretical investigations have not been done yet
on this material.

\section{Recent Experiments and Bose Gas Representation}

After a relatively dormant period of mid-90's, 
active researches of the field-induced ordering were
restarted on copper based dimer compounds.
The compound Cu$_2$(C$_5$H$_12$N$_2$)$_2$Cl$_4$, 
in which Cu$^{2+}$ ions form weakly-coupled ladders,
was investigated by various measurements;
the magnetization,\cite{ChaboussantETAL1997}
the ac susceptibility,\cite{HammerRBT1998}
the heat capacity\cite{HammerRBT1998} and
the inelastic neutron scattering.\cite{HammerRBT1998}
The heat capacity measurement at higher magnetic field
was performed later.\cite{HagiwaraKSM2000}
The sharp cusp was observed at the transition temperature
and a parabolic curve of the critical temperature
as a function of the magnetic field was drawn.

Several other materials were also investigated.
A heat capacity measurement showed a clear
anomaly in (CH$_3$)$_2$CHNH$_3$CuCl$_3$.\cite{ManakaYHKK1998}
An $S=1$ dimer compound Ba$_3$Mn$_2$O$_8$,
in which Mn$^{5+}$ ions carry $S=1$ spins and form singlet dimers,
was studied\cite{UchidaTBG2001} 
to reveal two step magnetization curve,
corresponding to the three possible values of
the dimer magnetization.

Even more thoroughly investigated was the family of 
compounds XCuCl$_3$ where X $=$ K, Tl, NH$_4$, etc.
From susceptibility measurments\cite{TakatsuST1997}
the broad peak temperature $T_p$ was obtained for
KCuCl$_3$ and TlCuCl$_3$ at zero field.
For these two compounds, the magnetization process
was investigated at finite external field,
\cite{ShiramuraETAL1997}
which yielded estimates of the critical magnetic 
field and the gap.
An inelastic neutron scattering was carried out
for \KCC\cite{CavadiniETAL1999} and later for
\TCC\cite{OosawaETAL2002a}, which showed
a three-line structure in the magnon excitation spectrum
corresponding to the three triplet excitations.\cite{CavadiniETAL1999} 
By comparing the experimentally determined dispersion
with the RPA approximation results, the exchange couplings
were also estimated.\cite{CavadiniETAL1999}

The experimental evidences were accumulated and 
it was realized that, apart from the correct critical 
exponents for the phase transition, there are several 
additional qualitative features that could not be 
accounted for by the mean field theory;\cite{TachikiY1970a,TachikiY1970b}
(i) the upward convex curvature in the adiabatic magnetization curve,
(ii) the upward convex curvature in the magnetization curve as a
function of the temperature with fixed magnetic field, and
(iii) the asymptotic behavior of the critical magnetic field
as a function of the temperature that seems more like
an algebraic dependence than an exponential dependence.

The upward convex curvature in the adiabatic magnetization had
been known from the very early period of the study.
\cite{HasedaETAL1970}
The temperature dependence of the magnetization was 
studied for \TCC\cite{OosawaIT1999}
and, at sufficiently strong fixed magnetic fields,
revealed an upward convex curvature and a dip at a 
temperature slightly higher than the transition point.
The critical magnetic field $H_{c1}(T)$ was estimated
for various temperature and the fitting to
\begin{equation}
  H_{c1}(T) - H_{c1}(0) \propto T^{\phi}
  \label{eq:AlgebraicDependence}
\end{equation}
yielded $\phi = 1.7(1)$.\cite{OosawaIT1999}
For the estimate of $\phi$, there are variations:
$\phi = 2.2(1)$,\cite{NikuniOOT2000}
$\phi = 2.0(1)$,\cite{TanakaETAL2001}
$\phi = 2.3(1)$\cite{OosawaETAL2002b} and
$\phi = 1.67(7)$\cite{ShindoT2004}
for \TCC; and 
$\phi = 2.6(2)$\cite{PaduanFilhoGO2004a,PaduanFilhoGO2004b}
for NiCl$_2\cdot$4SC(NH$_2$)$_2$.
It is notable that all the estimates so far are higher 
than the mean-field prediction $\phi = 3/2$ 
as discussed below.

The algebraic dependence \Eq{AlgebraicDependence}
is in contrast to the mean-field theory
which indicate an essential singular behavior near zero temperature:
$
  H_{c1}(T) - H_{c1}(0) \propto \mbox{exp}(-T_0/T)
$
with $T_0$ being a constant.
At least this inconsistency can be removed by introducing
a mean-field theory (the HF approximation)
based on the bosonic representation\cite{NikuniOOT2000}
of the Hamiltonian.
More importantly perhaps,
it provides a simple picture for the mechanism of the
field-induced magnetic ordering; the BEC of triplet 
quasiparticles.
This theory is equivalent to the spin wave theory with the
$M_z=0$ state taken as the vacuum state.
A similar spin wave theory was considered in an early study.
\cite{TachikiYM1970a}
It was found that the spin wave theory yields
properties closer to the experimental observations
than the mean-field theory in the spin representation.
For example, the spin wave theory correctly yields
the upward convex adiabatic magnetization curve 
in the ordered region,
whereas the mean-field theory yields a flat straight line.
Another important feature that the spin wave theory predicted
\cite{TachikiYM1970a}
was the the linear dispersion of the magnon in the ordered state.
However, the critical behavior was not discussed extensively
in early studies.
The HF approximation with the bosonic representation predicts
that the phase boundary in the $T-H$ plane has an algebraic
singularity near zero temperature with the
exponent $\phi=3/2$, in contrast to the experimental estimates.
It was reported\cite{ShermanLBOT2003} that by assuming
a relativistic form for the dispersion relation of magnons
the phase boundary can be made closer to those estimated
experimentally in an intermediate temperature region,
while the asymptotic behavior in the zero-temperature limit 
remains the same.
A similar observation was made in an improved HF calculation
\bcite{MisguichO2004} based on a realistic dispersion 
relation determined by experiments.

The BEC mechanism was confirmed by an
inelastic neutron scattering experiment.\cite{RueggETAL2003}
A linear dispersion of magnons (i.e., tripletons)
was observed in the ordered phase, 
as predicted by the spin wave theory\cite{TachikiYM1970a} 
and later also by bond-operator formalism.\cite{MastumotoNRS2002}
This was the first experimental observation of the magnon dispersion
in the ordered phase, while a similar preceding experiment
\cite{CavadiniETAL1999} was carried out in the
disordered phase.
In the language of Boson problems, the linear 
dispersion can be interpreted as 
a Goldstone mode characteristic to the BEC.
This piece of evidence was taken as a clear indication of the
BEC nature of the phase transition.

a first-order phase transition
at the condensation point in contrast to the standard
scenario expected for the BEC.
The reason for this apparent disagreement 
has not been clarified.

Even without the magnetic field,
the BEC phenomena can be seen when a sufficiently strong
pressure is imposed.
An elastic neutron scattering experiment\cite{OosawaFOKT2003}
demonstrated the existence of the pressure-induced magnetic ordering.
The essential mechanism seems to be the same as that of the
field-induced transition; the competition between the gap and
the inter-dimer interaction.
Therefore it is expected that the resulting $H-p$ phase diagram 
at $T=0$ is simple and consists of two phases; the disordered
phase including the origin $H=p=0$ and the ordered phase surrounding
the ordered region.
However, the phase diagram obtained
by the magnetization measurement\cite{TanakaGFOU2003}
seems to consist of three phases,
though the very existence of the extra phase
or its identity has not been established yet.

While most of the compounds studied in conjunction with the
field-induced magnetic ordering are quasi one-dimensional systems,
the compounds BaCuSi$_2$O$_6$ offers an example of a quasi 
two-dimensional system.
An inelastic neutron scattering experiment was carried out
\cite{SasagoUZS1997} to reveal the two-dimensional nature 
in the magnon dispersion.
Recent heat capacity and magnetization measurements
\cite{JaimeETAL2004} clearly showed the lambda peak in the
heat capacity.

The effect of disorder is certainly an interesting subject to explore. 
It has been already studied through the magnetization measurement
\cite{OosawaT2002,ShindoT2004}
of the compound (Tl$_{1-x}$K$_x$)CuCl$_3$.
At a finite temperature, the presense of the disorder showed
only a quantitative effect, namely, reduction of the transition 
temperature\cite{OosawaT2002} and did not alter the nature 
of the phase transition in a qualitative way.
Qualitative differences could be observed, however, as the 
temperature was lowered.
For instance, the asymptotic behavior of the critical magnetic
field seemed to deviate from that of the homogeneous case.
Namely, the value of the exponent $\phi$ appeared to change
when the disorder was introduced, as suggested in the
scaling theory.\cite{FisherWGF1989}

Another type of disorder, namely non-magnetic impurities,
was studied for Tl(Cu$_{1-x}$Mg$_x$)Cl$_3$.
\cite{OosawaOT2002,OosawaFKT2003}
Upon Mg doping, a magnetic phase transition is induced even at
zero magnetic field due to the presense of uncoupled
magnetic ions.
It was found that this impurity-induced magnetic ordering
does not affect the dispersion of the magnons at zero magnetic field.
However, experiments in the magnetic field are yet to be performed
in order to observe the effect upon the critical phenomena
at the QCP.

\section{Scaling and Numerical Results}

As stated above, the bosonic theory
explains most of the characteristics observed in experiments.
However, it still has a short-coming, i.e.,
it is destined to produce wrong critical properties
for the phase transition at finite temperature
and there is no systematic way of amending the flaw.
For example, it results in a discontinuous drop 
in the magnetization at the critical point 
as we increase the temperature,
whereas we only see a dip in experiments.
\cite{OosawaIT1999}
Another apparent disagreement is of the estimate
of the exponent $\phi$.
Experiments seemed to indicate a larger 
value for the exponent $\phi$ for the QCP
than the HF theory does.
Since the incorrectness of the mean-field type theory 
concerning the critical exponent is well-known,
the disagreement seems trivial at a first glance.
However, it is not trivial and in fact the HF theory may be
correct concerning the QCP as we see below.

The first quantum Monte Carlo simulation
for clarifying the critical properties\cite{WesselOH2001}
was performed for the model Hamiltonian \Eq{OriginalHamiltonian}
representing weakly coupled ladders.
The estimates of the critical exponent $\phi$ differed
from the HF value; $\phi = 1.8(2)$ for the upper critical field
and $\phi = 3.1(2)$ for the lower critical field.
Another Monte Carlo simulation was carried out
\cite{NohadaniWNH2004} for a similar system.
A temperature-dependent effective exponent $\phi(T)$
was defined so that it characterizes the $H_c-T$ curve in 
a finite temperature-range centered at $T$.
Again, $\phi(T)$ turned out to be greater than 3/2.
However, it decreases as the temperature is lowered,
and it was concluded that the trend is consistent with
the HF value.

Recently, we carried out another Monte Carlo simulation
\cite{Kawashima2004}
to further clarify the problem concerning the critical 
properties at QCP.
Since the critical properties of the effective Hamiltonian
\Eq{ReducedHamiltonian} 
derived with the Tachiki and Yamada's mapping
is expected to be the same as those of
the original Hamiltonian \Eq{OriginalHamiltonian},
we chose to work on the effective Hamiltonian for simplicity.
The effective spin Hamiltonian is a uniform $S=1/2$ $XXZ$ model 
in three dimensions with a spatial anisotropy 
and with the easy-plane spin anisotropy.
Since the spatial anisotropy and the $S^z-S^z$ couplings
should not affect the critical properties, 
we can drop them without changing critical properties.
Therefore, we arrive at the $XY$ model Hamiltonian.
\begin{equation}
  \Ham =
         - J \sum_{(ij)}
             \left( S_i^x S_j^x + S_i^y S_j^y  \right)
         - H \sum_i S_i^z.
  \label{eq:XYHamiltonian}
\end{equation}

The simulation method employed is based on
the directed-loop algorithm \bcite{SyljuasenS},
for which a review can be found in
an recent article.\bcite{KawashimaH2004}
The magnetization at zero temperature was computed as a function of
the magnetic field.
It was found that in the thermodynamic limit the deviation of the
magnetization from the saturation value depends 
linearly on the deviation of the magnetic field from the critical value:
$1/2-\langle S_i^z \rangle$ $\propto H_{\rm c1}-H$
where $H < H_{\rm c1}$.
This is consistent with the HF prediction and also with the experiments.

For the susceptibility of the perpendicular components of spins,
$\chi \equiv \sum_{\vect r}\int_0^{\beta} d\tau
\langle S^x(\vect r, \tau) S^x(\vect 0, 0) \rangle$,
the result suggested that the system size dependence
at the critical point $H=3J$ was characterized by
$\chi(\beta/L^2,H=H_{\rm c}) \propto L^{2.5}$.
As we see below, this is also consistent with 
the scaling theory that produces the mean-field type 
scaling properties.

The temperature dependent critical field $H_c(T)$
was also computed.
A set of values of temperature were chosen and 
for each temperature the susceptibility was computed as a function
of the magnetic field.
We then carried out the finite-size scaling analysis
with a fixed temperature to obtain the critical field 
at the temperature.
The finite size scaling plot worked nicely with the critical exponents
estimated for the classical $XY$ model, namely,
$\nu = 0.67155(27)$ and $\eta = 0.0380(4)$.\cite{CampostriniHPRV2001}
In this fashion, we estimated the critical field at 
various temperatures.
A remark may be useful here
concerning the critical properties at a finite temperature.
There has been a confusion about the critical behavior of 
the specific heat, which was considered to be divergent in some 
early articles.
In fact, the specific heat exponent $\alpha = -0.0146(8)$
\cite{CampostriniHPRV2001} is negative, i.e., it has a sharp
cusp but finite even at the critical point.

In \Fig{PhaseDiagram}, two curves are plotted for comparison.
The solid curve represents the mean-field critical exponent,
$\phi = 1.5$, whereas the dashed curve the previous estimate
$\phi = 2.0$\cite{TanakaETAL2001} based on an experiment 
on TlCuCl$_3$.
At a first glance, it seems that the curve with $\phi=2.0$
fits the data better than $\phi=1.5$.
However, when the logarithmic scale is used, as we do in
the inset of \Fig{PhaseDiagram}, 
it is rather clear that $\phi=2.0$
explains only a transient behavior, and the correct
asymptotic value of the exponent is $\phi=1.5$.

\deffig{PhaseDiagram}{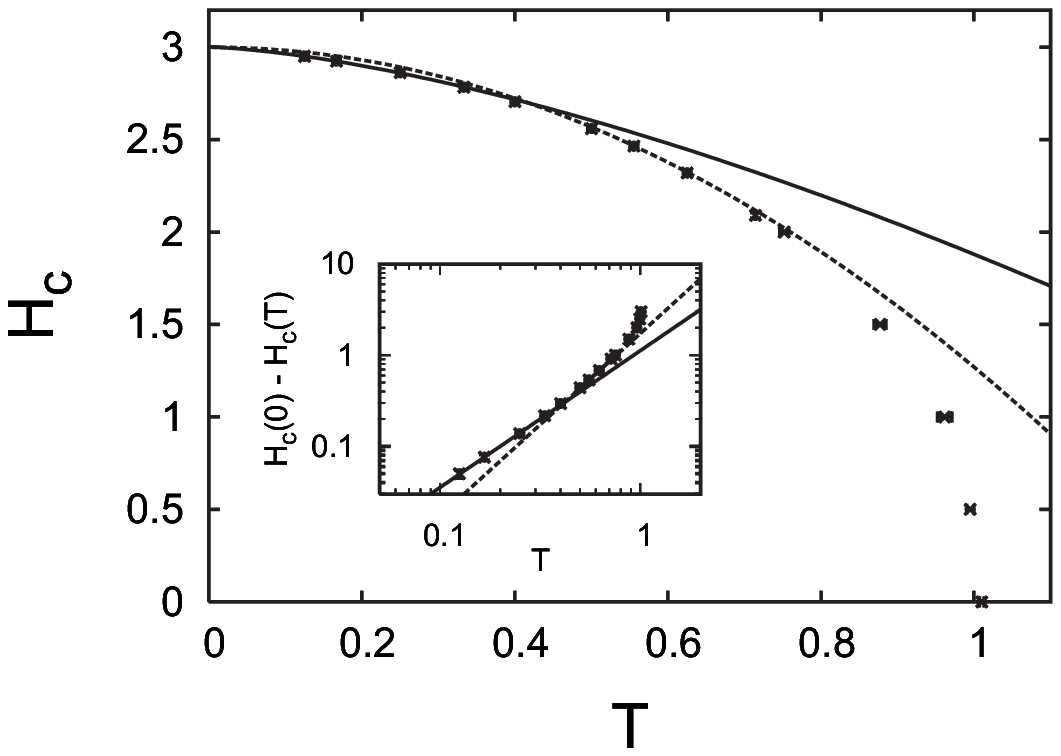}{0.45}{
  Critical field as a function of temperature.
  The solid curve corresponds to $\phi= 1.5$ whereas
  the dashed one $\phi=2.0$.
  The inset is the same data plotted in the logarithmic scale.
  While in the intermediate temperature region $\phi=2.0$ seem
  to fit the data well, in the low-temperature region, 
  the correct slope $\phi = 1.5$ yields a better fitting.
}

Now, let us see why the mean-field type scaling should hold
for the QCP in the present system.
To this end, we first derive the bosonic Hamiltonian
from \Eq{XYHamiltonian} by using the mapping
$S_i^+ = b^{\dagger}_i$, $S_i^- = b_i$, and 
$S_i^z = \hat n_i - \frac12$
with the constraint
\begin{equation}
  \hat n_i \equiv b^{\dagger}_i b_i = 0, 1.
  \label{eq:Condition}
\end{equation}
Then we obtain
\begin{equation}
  \Ham =
         - t \sum_{(ij)}
             \left( b_i^{\dagger} b_j + b_j^{\dagger} b_i  \right)
         - \mu \sum_i \hat n_i
         + \Lambda \sum_i \hat n_i (\hat n_i - 1),
  \label{eq:BosonicHamiltonian}
\end{equation}
with new parameters $t$ and $\mu$ corresponding to the old ones 
as $t \sim J/2$ and $\mu \sim H$.
The $\Lambda$-term imposes the condition \Eq{Condition}, approximately.
However, we believe this approximation does not affect 
the critical properties since the difference between the hard
(original) constraint and the soft one is significant only when
more than double occupancies occur frequently, for which
the average occupation number needs to be large.
As we approach the QCP, however,
we have smaller particle density
and in the end we have zero density at the QCP.
Therefore, we expect the softening of the constraint has 
no effect on the critical properties.

The rest of the argument is essentially based on 
the results obtained earlier\cite{Uzunov1981,FisherWGF1989} 
for diluted bose gas.
Similar results can be found also in textbooks.\cite{Text}
However, the connection to the present model has
not been very transparent, which we can see in the fact that
those early scaling theories have been seldom mentioned 
in conjunction with the field-induced ordering in dimer systems.
Therefore we reproduce the scaling results here,
and it would be useful to have explicit formulas in the form 
directly comparable with the Monte Carlo results.
(In particular, explicit formulas for the system-size 
dependence are hardly found in the literature.)
First, the continuous field theory for the model
\Eq{BosonicHamiltonian} is characterized by the action
\begin{eqnarray}
  S & = & \int d\vect{x} \int d\tau\ 
         \left( \rule{0mm}{5mm}
             \psi^*\frac{\partial \psi}{\partial \tau}
           + \left|\nabla\psi\right|^2
         \right. \nonumber \\
    &   &
         \left. \rule{0mm}{5mm}
         \qquad
         - h\,{\rm Re}\,\psi 
         - r \left|\psi\right|^2
         + u \left|\psi\right|^4 
         \right)
         .
         \label{eq:Action}
\end{eqnarray}
Obviously, the fixed point is given by $h=r=u=0$.
A simple dimensional analysis suffices to obtain the
scaling dimensions at this fixed point.
The effect of the scale transformation around the fixed point
up to the scale $b$ is the following:
\begin{eqnarray}
  & &
      \beta \rightarrow \tilde \beta \equiv  \beta b^{-2}, \ 
      L \rightarrow \tilde L \equiv L b^{-1}, \
      \psi \rightarrow \tilde \psi \equiv \psi b^{\frac{d}{2}}, 
      \nonumber \\ 
  & &
      h \rightarrow \tilde h \equiv h b^{2+\frac{d}{2}}, \
      r \rightarrow \tilde r \equiv r b^2,\ 
      u \rightarrow \tilde u \equiv u b^{2-d}, \nonumber
\end{eqnarray}
where $L$ is the system size.
Note that the non-linearity represented by parameter $u$
diminishes as we renormalize for $d>2$.
Hence, the upper critical dimension $d_c=2$,\cite{Uzunov1981}
and the critical properties controlled by the
$u=0$ fixed point when $d > 2$.
For three dimensions, therefore, we should expect a
mean-field type scaling.
To be more concrete, let us derive the scaling properties 
for the magnetization, $m \sim \langle |\psi| \rangle \sim 
\langle S^x_i \rangle$.
We assume that a scaling property holds for the singular part of 
$\Phi \equiv -\log Z(h,r,u,\beta,L)$, namely,
$
  \Phi(h,r,u,\beta,L) 
  =
  \Phi(\tilde h, \tilde r, \tilde u, \tilde \beta, \tilde L)
$
for an arbitrary $b$. 

In order to obtain the asymptotic behavior
of the critical magnetic field, we note that
we can make $\tilde u$ small by choosing large $b$
so that the perturbation theory in $\tilde u$ is nearly exact 
for the renormalized system.
The perturbation theory yields that $\Phi$ for the
renormalized system has a 
singularity when the condition\cite{Text}
$
  \tilde r = \tilde r_c \equiv A \tilde u / \tilde \beta^{d/2}
$
is satisfied with $A$ being some numerical constant.
It means that in terms of the bare coupling constants
the singularity occurs when\cite{Uzunov1981}
$$
  H_c(T)-H_c(0) \propto T^{d/2}.
$$
This is also confirmed by the simulation.

\section{Summary}

We have presented a brief review of the field-induced magnetic ordering.
From the early stage of the research,
the existence of the ordering was established, 
which was also supported by the mean-field theory.
The other mean-field theory, i.e, the HF theory 
was later constructed based on the bosonic representation 
and described various features 
observed in the experiments more precisely.
Many of recent experimental results, however, seemed to contradict
even the HF theory in one important point, i.e., the critical 
property at the QCP.
There could be some possible sources of this discrepancy,
such as a presence of Dzyaloshinsky-Moria interactions,
a staggered component of the $g$-tensor giving rise to an effective
staggered field, or a crystal field that reduces the symmetry.
However, the recent Monte Carlo simulation results present
an explanation, though not the unique possibility, that 
the true asymptotic behavior is consistent with the scaling theory
while the transient behavior is hard to get rid of.

Interesting remaining problems concerning the field-induced 
magnetic ordering are 
(i) experimental and numerical investigation of the complete 
    phase diagram of a disordered magnet that has a finite gap
    in the pure case,
    and the critical phenomena at the bose-glass to 
    superfluid transition, and
(ii) the effect of a symmetry lowered by the other
    kinds of interactions (crystal fields, effective 
    staggered field, etc) or, in case where the
    gap is caused by the easy-plane axial anisotropy,
    by a transverse field tangent to the easy plane.

The author thanks M.~Oshikawa and H.~Otsuka for helpful comments.
The present work was supported by the grant-in-aid (Program No.\ 14540361)
from Monka-sho, Japan. 
Our computation presented in this article was performed 
using the computers at Supercomputer Center, 
Institute of Solid State Physics, University of Tokyo.


\end{document}